\RequirePackage{amsthm}
\documentclass[default,iicol]{sn-jnl}

\usepackage{booktabs}
\usepackage{graphicx}%
\usepackage{multirow}%
\usepackage{amsthm,amssymb,amsfonts}%
\usepackage{amsfonts}
\usepackage{amsmath}%
\usepackage{mathrsfs}%
\usepackage[title]{appendix}%
\usepackage{xcolor}%
\usepackage{textcomp}%
\usepackage{manyfoot}%
\usepackage{booktabs}%
\usepackage{bm}
\usepackage{algorithm}%
\usepackage{algorithmicx}%
\usepackage{algpseudocode}%
\usepackage{listings}%
\usepackage{natbib}

\renewcommand{\bf}{\mathbf{f}}

\usepackage{natbib}
\usepackage{soul}

\theoremstyle{thmstyleone}%
\newtheorem{theorem}{Theorem}%  meant for continuous numbers
\newtheorem{proposition}[theorem]{Proposition}% 

\begin{document}

\title[Article Title]{Roughness regularization for functional data analysis with free knots spline estimation }
\author*[1]{\fnm{Anna}\sur{De Magistris}}\email{anna.demagistris@unicampania.it}
\author[1]{\fnm{Valentina}\sur{De Simone}}\email{valentina.desimone@unicampania.it}

\equalcont{These authors contributed equally to this work.}
\author*[1]{\fnm{Elvira} \sur{Romano}}\email{elvira.romano@unicampania.it}
\equalcont{These authors contributed equally to this work.}

\author[1]{\fnm{Gerardo} \sur{Toraldo}}\email{gerardo.toraldo@unicampania.it}
\equalcont{These authors contributed equally to this work.}

\affil*[1]{\orgdiv{Dipartimento di Matematica e Fisica, University of Camapania "Luigi Vanvitelli"}, \orgaddress{\street{Viale Abramo Lincoln 5}, \city{Caserta}, \postcode{81100}, \state{Italy }, \country{Caserta}}}

\abstract{In the era of big data, an ever-growing volume of information is recorded, either continuously over time or sporadically, at distinct time intervals. Functional Data Analysis (FDA)  stands at the cutting edge of this data revolution, offering a powerful framework for handling and extracting meaningful insights from such complex datasets. The currently proposed FDA me\-thods can often encounter challenges, especially when dealing with curves of varying shapes. This can largely be attributed to the method's strong dependence on data approximation as a key aspect of the analysis process. In this work, we propose a free knots spline estimation method for functional data with two penalty terms and demonstrate its performance by comparing the results of several clustering methods on simulated and real data.}


\maketitle
\section{Introduction}\label{sec1}

In recent years, there has been a rapid increase in the availability of large data resulting from the observation of a phenomenon along a continuous domain, such as time, space, or frequency domains. The growing need to analyze data of this type, characterized by an intrinsic  functional nature, is the basis for the development of the field of Functional Data Analysis (FDA) \cite{bib17},\cite{bibr}. 
Functional data analysis typically involves two key steps: first, representing discrete observations as functions and then applying statistical methods. The step of functional representation helps to reduce noise and fluctuations in the data and plays a key rule in the FDA because it sets the foundation for subsequent statistical analyses. There are several methods and techniques that can be employed to perform functional representation. Some common approaches include smoothing spline, wavelet analysis, Fourier analysis, principal component analysis, smoothing techniques, and interpolation methods.
 The choice of the functional representation method depends on the nature of the data, the specific goals of the analysis, and the assumptions made about the underlying structure of the functions. Each method has its advantages and limitations, and the selection should be based on the characteristics of the data and the objectives of the analysis.
To address the efficient representation and analysis of functional data with various shapes, we focus on the problem of representing functional data using free knots spline estimation. In the last years, the problem of representing functional data with free knots spline has been addressed by the same authors. Gervini \cite{bib2} proposes free knots regression spline estimators for the mean and the variance components of a sample of curves, showing that free knots splines estimate salient features of the functions (such as sharp peaks) more accurately than smoothing splines.
Inspired by a Bayesian model \cite{bib20} propose estimating various curves using piecewise polynomials. The first one establishes a joint distribution over the number and positions of knots defining the piecewise polynomials and employs reversible jump Markov Chain Monte Carlo methods for posterior computation. The second, based on a marginalised chain on the knot number and locations, provides a method for inference about the regression coefficients and functions of them in both normal and non normal models.
Among others, a novel knot selection algorithm is introduced by \cite{bib17} for nonparametric regression using regression splines.
Recently \cite{bib16},  building upon a previous proposal of \cite{bib18}, introduces a data-driven approach that uses a machine learning-style algorithm to select knots efficiently and employs orthogonal spline bases (splinets) to represent sparse functional data. 
In this work, generalizing 'the balanced-discrepancy principle' proposed by \cite{bib22} in the functional data analysis framework, we propose a simple double penalization criterion to improve the smoothing process with free knots spline approximation for various types of curve shapes. This approach is designed to improve the precision and overall effectiveness of fitting spline curves to diverse datasets. The paper is organized as follows. Section $2$ introduces mathematical foundations of smoothing with roughness penalty. In section $3$, we illustrate free knots spline estimation and their construction with a double penalization criterion. In section $4$ we illustrate performance of the proposed penalization criterion on simulated data. An extensive comparative analysis to evaluate which approximation method is most suitable for functional data in terms of clustering performance is conducted. Section $5$ shows the performance of a clustering method on COVID-19 pandemic data for 30 countries by comparing free knots spline and free knots spline with double penalization criterion. 

\section{Smoothing Functional Data}\label{sec2}

Let $\{x_i(t), \; i = 1, . . . , n\}$  be a sample of real-valued functions on a compact interval  $T$ related to a functional variable $X$.
The sample curves can be considered realizations  of a stochastic process $X = \{X(t) : t \in T \}$ whose sample functions belong to the Hilbert space
$L^2(T )$ of square integrable functions with the usual inner product $f,g = \int_T f(s) g (s) ds$.

In real applications, $x_i(t)$ often cannot be observed directly, but may be collected discretely over time points $\{t_1,...,t_h\} \subset T$.
FDA aims  to reconstruct the true functions from the observed data
using  the basis function approach, that reduces the estimation of the function $x_i(t)$ to the estimation of a finite linear combination
of known basis functions
$\phi_k(t), \; k=1 \hdots n_\mathcal{B}$
\begin{equation*} 
     {x}_j(t) = \sum_{k=1}^{n_{\mathcal{B}}} \textbf{c}_{k}^j {\phi}_k({t}), \: \: \: j = 1,...,n
\end{equation*}
where $\mathbf{c}^j \in \mathbb{R}^{n_{\mathcal{B}}}$ is a vector of coefficients.
The choice of the base and the dimension $n_{\mathcal{B}}$ depend strongly on the nature of the data. The most known bases are: the Fourier base, B-spline, polynomial, exponential, wavelets. The Fourier base is used when the data has a cyclical nature, the exponential base when the data shows exponential growth, and the B-spline base are the most used when the data does not have a strong cyclical trend.
There are different ways of obtaining the basis coefficients depending on the kind of observations.  Generally  observed data are contaminated by random noise that can be viewed as random  fluctuations around a smooth trajectory,
or as actual errors in the measurements. %The observed data are separated into two components are %

The observed data is collected in a matrix $\textbf{Y} \in \mathbb{R} ^ {h \times n}$ whose elements are:
\[ y_{i,j}={x}_j({t}_i)+ \boldsymbol{\epsilon}_{i}^j, \;\;\; i=1,...,h;\;  j=1,...,n\]
where $\boldsymbol{\epsilon}_{i}$
is unobserved  error term independent and identically distributed random variable
%where is assumed to be an unobservant error term with distributed independently and identically 
with zero mean and constant variance
\( \sigma^2 \).
In this case an appropriate way to estimate the basis coefficients from the data 
%is a smoothing technique as the least square one:
is by using a least squares approximation
\[ \hat{\textbf{C}} = \underset{{\textbf{C}}}{\operatorname{argmin}} \left\| \textbf{Y} -\boldsymbol{\Phi}^T \textbf{C}\right\|^2_2
,
%\sum_{j=1}^\textbf{y} [y_j - \sum_{k}^{{n_{\mathcal{B}}}}c_k  \phi_k(t_j)]^2, 
\]
with  \( \boldsymbol{\Phi}  \in \mathbb{R}^{{n_{\mathcal{B}} \times h }}\),
%, with $rank(\Phi)={N_{\mathcal{B}}}$,
$\boldsymbol{C} \in \mathbb{R}^{  {n_{\mathcal{B}} } \times n}$, where $\phi_{ij}=\boldsymbol{\phi}_k(\mathbf{t}_j)$ and $c_{ij}=\mathbf{c}_j^{i}$.

The smoothness is implicitly controlled by the number of basis functions, ${n_{\mathcal{B}}}$.
If we assume that $n_{\mathcal{B}} \leq h$ and $rank(\boldsymbol{\Phi})=n_{\mathcal{B}}$ then

\begin{equation}
\hat{\textbf{C}} = {(\boldsymbol{\Phi}\boldsymbol{\Phi}^T)}^{-1} \boldsymbol{\Phi}\textbf{Y}.
\end{equation}
and  the matrix \( \hat{ \textbf{Y}} \) of the approximation values is:
                       
\begin{equation}
\hat{\textbf{Y}} = \boldsymbol{\Phi}^{T} \hat{\textbf{C}} = \boldsymbol{\Phi}^{T} {(\boldsymbol{\Phi}\boldsymbol{\Phi}^T)}^{-1} \boldsymbol{\Phi}\textbf{Y}.
\end{equation}

Increasing ${n_{\mathcal{B}}}$ leads to overfitting and generates a \(x\) curve that is overly "wiggly. One way to overcome this
drawback is to use a roughness penalty term.

\subsection{Smoothing functional data with a roughness penalty}\label{subsec2}

The roughness penalty approach defines a measure of the roughness of the fitted function $x$ using the derivatives of some order $m \geq 1$
\begin{equation}
    PEN_m(x) = \int {[D^mx(s)]}^2 ds.
\end{equation}
This allows to measure the closeness of $x(t)$ to a polynomial of order $m$. The most commonly used roughness penalty is $m=2$ that permits to keep under control the curvature of the curve, that is we control the variability of the slope of the curve \cite{bib4}.  

Often, there is a need for a wider class of measures of deviation. Especially,
when there is periodicity in the data or an exponential trend, it would not be
sufficient to use the integrated squared mth derivative because it can only penalize
deviations from polynomials.
More generally, a measure of roughness is then given by
\begin{equation}
    PEN_L(x) = \int {[Lx(s)]}^2 ds,
\end{equation} 
where $L$ is the linear differential operator defined as \cite{bibr}
\begin{align*}
Lx(t)=\alpha_0(t) x(t)+\alpha_1(t) D^{1}x(t)+....+ \\ + \alpha_{m-1}(t) D^{
m-1} x (t)+ D^{m}x(t).
\end{align*}
%{\small\[Lx(t)=\alpha_0(t) x(t)+\alpha_1(t) D^{1}x(t)+....+\alpha_{m-1}(t) D^{
%m-1} x (t)+ D^{m}x(t).\]}
Obviously $PEN_m(x)$ is a special case of $PEN_L(x)$ with $\alpha_0(t)=\alpha_1(t)=...=\alpha_{m-1}(t)=0$ and $\alpha_m(t)=1$. In the following we will assume the $\alpha_i(t)$ to be constant.
Then, the penalized least squares approximation is given by

\begin{equation}
     \hat{\textbf{C}} = \underset{{\textbf{C}}}{\operatorname{argmin}} \left\| \textbf{Y} -\boldsymbol{\Phi}^T \textbf{C}\right\|^2_2 + \lambda \textbf{C}^T \mathbf{R}_L \textbf{C},
\end{equation}
%In this case highly variable functions are penalized 
%because their curvature can be have  large values.
with  $\mathbf{R}_L=\alpha_0 \mathbf{R}_{0}+\alpha_1 \mathbf{R}_{1}+...\alpha_m \mathbf{R}_{m}$ discretization of $PEN_L(x)$, where
$ \mathbf{R}_{l}
\in \mathbb{R}^{{n_{\mathcal{B}}} \times {n_{\mathcal{B}}}}$ ($0\leq l\leq m)$  :
\begin{equation}
    (r_{l})_{ij} = \int_{I} D^l \phi_i(s) D^l \phi_j(s)^T ds, 
\end{equation}
where $I$ is a suitable interval containing the data.
The regularization parameter 
 \( \lambda >0\)  manages the compromise between the fitting to the data and the roughness of the  function: the smaller it is \( \lambda \), the closer the estimate is to the estimate of least squares and tends to interpolate the observed points; the higher  is \( \lambda \), the flatter the smooth function tends to be. This parameter can be calibrated through a generalised cross validation (GCV)\cite{bibr}.
%For using a $PEN_L$ its mth derivatives  and all functions $\alpha_j(t), \; \; j = 0, . . . ,m- 1$ must be known, then the computational cost can be prohibitive.
In this paper we consider the first and second-order roughness in order to control  both the course and the variability of slope of the curve and to avoid placing an excessive burden on the cost of the penalty matrix. Then the
penalized least square problem becomes

\begin{align}
\label{solution}
 \hat{\textbf{C}} = \underset{{\textbf{C}}}{\operatorname{argmin}} \left\| \textbf{Y} -\boldsymbol{\Phi}^T \textbf{C}\right\|^2_2 + \lambda_1 \textbf{C}^T \mathbf{R}_1 \textbf{C} + \\ + \lambda_2 \textbf{C}^T \mathbf{R}_2 \textbf{C} \ \ \ \ \ \ \ \lambda_2>0, \lambda_1>0.
 \notag
\end{align}

The solution of  \eqref{solution} can be computed by solving the system arising by first order optimality conditions
\begin{equation}
    \hat{\textbf{C}} = {(\boldsymbol{\Phi} \boldsymbol{\Phi}^T + \lambda_1\textbf{R}_1 + \lambda_2 \textbf{R}_2)}^{-1} \boldsymbol{\Phi} \textbf{Y}.
\end{equation}
The expression for the data-fitting \(\hat{\textbf{Y}}\)  is: 

\begin{equation}
\hat{\textbf{Y}}  = \boldsymbol{\Phi}^T{(\boldsymbol{\Phi} \boldsymbol{\Phi}^T + \lambda_1\textbf{R}_1 + \lambda_2 \textbf{R}_2)}^{-1} \boldsymbol{\Phi} \textbf{Y}
\end{equation}

\section{Free knots spline }\label{sec3}

For their compact support and fast computation, as well as the ability to create smooth approximations of  non periodic data, B-splines are a common choice in the functional data framework. 
The approximations by splines can be significantly improved if knots are allowed to be free rather than
at a priori fixed locations \cite{bib2}.

It is well-known that the primary advantage of free knots spline over smoothing splines is their greater flexibility in modelling data, as they allow for a better adaptation of the curve's shape to the specific characteristics of the data.

The free knots spline is a spline in which the positions of the knots are considered parameters to be estimated by the data. Adjusting the position of the nodes allows you to adapt the shape of the function to the target function.

This section considers smooth estimators for knots selection for the approximation of a given curve. Existing approaches are based on individual levelling of sample curves, followed by the mean of the cross-section and the calculation of covariance \cite{bib5}. These methods, however, do not take strength from the dataset that we have available in the leveling phase. A further drawback is that analytical expressions for optimal knots locations, or even for the general characteristics of optimal knots distributions, are not easy to derive.
We introduce free knots spline estimators that avoid individual levelling. We show that this approach applied to the methods seen in the previous section (smoothing spline with one parameter and with two parameters) often produces better estimators than sanding splines \cite{bibr} in which knots are chosen randomly and equally at the cost of a modest increase in computational complexity. In this section we introduce the algorithm of optimal knots placement. Given a vector of nodes  $\boldsymbol{\tau} \in \Re^p$ with
$a<\tau_1<\tau_2..<\tau_p<b$, the Jupp\cite{bib6} transformation  
$\textbf{k}$ of
$\boldsymbol{\tau}$, $ \textbf{k}=J( \boldsymbol{\tau})$, is defined componentwise as 
\begin{equation*}
    {k}_i = \log\frac{(\boldsymbol{\tau}_{i+1} - \boldsymbol{\tau}_i)}{(\boldsymbol{\tau}_i - \boldsymbol{\tau}_{i-1})} \: \: \: \: \: \: \: \: \: i=1,...,p
\end{equation*} 
where \( {\tau}_0 = a\) and \({\tau}_{p+1} = b \). This one-to-one transformation maps constrained and ascending knots vectors \(\boldsymbol{\tau}\) on unconstrained and unclassified carriers \(\textbf{k}\), which has some practical and theoretical advantages.\\
Note that for each fixed set of knots, the class of such splines is a linear space of functions with $(p+r) = {n_{\mathcal{B}}}$ free parameters where $r$ is the order of spline and $p$ is the length of knots.
%, while if the locations of the knots are free variables has  $(2p + r)$ free linear and nonlinear parameters.

Let \(\boldsymbol{\phi}(\textbf{t},\textbf{k}) \in \mathbb{R}^{{n_{\mathcal{B}}}}\) be the vector of the basic functions B-spline corresponding to a set of nodes \(\textbf{k}\) and \(\boldsymbol{\Phi}(\textbf{k})\) the matrix \(h \times {n_{\mathcal{B}}} \) 
whose j-row is \( \boldsymbol{\phi}(\textbf{t}_j,\textbf{k})^T\).  We find the coefficients of linear expansion and the vector of the optimal nodes minimizing the penalized least squares problem\footnote{Note that for $\lambda_1=\lambda_2=0$ (\ref{P1}) becomes the classic free knots spline problem.}:

\small{
\begin{align}\label{P1}
(\hat{\textbf{C}},\hat{\textbf{k}}) = \underset{({\textbf{C}},{\textbf{k}}) \in \mathbb{R}^{n_{\mathcal{B}}}\times \mathbb{R}^p}{\operatorname{argmin}} \left\| \textbf{Y} -\boldsymbol{\Phi}^T\textbf{C}\right\|^2_2 + \lambda_1 \textbf{C}^T\textbf{R}_1 \textbf{C} + \\ + \lambda_2 \textbf{C}^T\textbf{R}_2 \textbf{C} \ \ \ \ \ \lambda_1 \ge 0, \lambda_2 \geq 0
\notag
\end{align}}

Alternating minimization algorithm can be used to solve problem (\ref{P1}); at each iteration the problem is splitted in two optimization ones.
Closed form solution can be obtained for minimization with respect to $\textbf{C}$ fixing  $\textbf{k}$.  The optimal value can be obtained by solving the system with coefficient matrix
\[ \mathbf{H}(\mathbf{k})= \boldsymbol{\Phi}(\textbf{k}) \boldsymbol{\Phi}(\textbf{k})^T + \lambda_1 \textbf{R}_1 + \lambda_2 \textbf{R}_2.\]

\begin{proposition}
If the $\boldsymbol{\Phi}(\textbf{k})$ is full rank, the matrix $\textbf{H(\textbf{k})}$ is SPD and let $\sigma(\textbf{H})$ be the set of eigenvalues of matrix $\textbf{H(\textbf{k})}$ we have
$\sigma(\textbf{H})\subseteq [\sigma^-, \sigma^+]$, with
\[\sigma^-=\sigma_{min}(\boldsymbol{\Phi}(\textbf{k}) \boldsymbol{\Phi}(\textbf{k})^T) + \lambda_1 \sigma_{min}(\textbf{R}_1)+ \lambda_2 \sigma_{min}(\textbf{R}_2)\]
\[\sigma^+=\sigma_{max}(\boldsymbol{\Phi}(\textbf{k}) \boldsymbol{\Phi}(\textbf{k})^T) + \lambda_1 \sigma_{max}(\textbf{R}_1)+ \lambda_2 \sigma_{max}(\textbf{R}_2)\].
\begin{proof}
$\textbf{H(\textbf{k})}$ is the sum of $\boldsymbol{\Phi}(\textbf{k}) \boldsymbol{\Phi}(\textbf{k})^T$  that is SPD, and of other matrices that are non negative definite \cite{bib4}; then the first statement follows. For the second statement we recall that if $\textbf{A}$ and $\textbf{B}$ are real, symmetric matrices, then $\textbf{A + B}$ has real eigenvalues, and the following inequalities hold

\begin{align*}
    \sigma_{min}(\textbf{A})+\sigma_{min}(\textbf{B}) \leq \sigma_{min}(\textbf{A + B}) \leq  \\ \leq \sigma_{max}(\textbf{A + B}) \leq \sigma_{max}(\textbf{A})+\sigma_{max}(\textbf{B}).
\end{align*}
%\[\sigma_{min}(\textbf{A})+\sigma_{min}(\textbf{B}) \leq \sigma_{min}(\textbf{A + B}) \leq  \sigma_{max}(\textbf{A + B}) \leq \sigma_{max}(\textbf{A})+\sigma_{max}(\textbf{B}).\]
\end{proof} 
\end{proposition}

Proposition 1 suggests to use  Cholesky factorization to compute

\begin{equation*}
    \hat{\textbf{C}}(\textbf{k}) = \mathbf{H}(\mathbf{k})^{-1} \boldsymbol{\Phi}(\textbf{k}) \textbf{Y}.
\end{equation*}

The minimization  with respect to $\textbf{k}$ fixing  $\textbf{C}$ gives the following nonlinear optimization problem:

%\begin{equation}\label{P2}
    %\hat{\textbf{k}} = \underset{\textbf{k} \in  \mathbb{R}^p}{\operatorname{argmin}} \sum^{n}_{i=1} \left\|\textbf{x}_i - B(\textbf{k}){\{B(\textbf{k})^T B(\textbf{k}) +\lambda_1 R_1+ \lambda_2 R_2\}}^{-1} B(\textbf{k})^T \Bar{x}\right\|^2
%\end{equation}
%{\color{red}
%{ nella 12 manca dipendenza da j}
%Nella propositio non è definiton $\sigma$. }
%{\color{blue} Risposta di Vale: sigma(H) è lo spettro di H... è definito nella Prep. La 12 definisce un funzione da $\Re^j$ a $\Re$, quindi j è la dimensione dello spazio}
\begin{equation}\label{P2}
    \hat{\textbf{k}} = \underset{\textbf{k} \in  \mathbb{R}^p}{\operatorname{argmin}}  \left\|\textbf{Y} - \boldsymbol{\Phi}(\textbf{k})^T\textbf{H}(\textbf{k})^{-1} \boldsymbol{\Phi}(\textbf{k})\textbf{Y}\right\|^2
\end{equation}

To solve (\ref{P2}), we apply the knots addition algorithm that produces knots sequences of increasing dimensions. We define the functional $f_j : \mathbb{R}^j \rightarrow \mathbb{R} $ as follows:

\begin{equation}\label{P3}
    f_j(\textbf{k}) = \left\|\textbf{Y} - \boldsymbol{\Phi}(\textbf{k})^T\textbf{H}(\textbf{k})^{-1} \boldsymbol{\Phi}(\textbf{k}) \textbf{Y}\right\|^2 \: \: \:  1 \leq j \leq p.
\end{equation}

Now, we will present the procedure of the gradual node addition algorithm\cite{bib2}.\\

\fbox{
\begin{minipage}{7cm}

\begin{center}
Gradual node addition algorithm    
\end{center} 

\vspace{0.2cm}
\textbf{Initialization}
\begin{itemize}
\item[] Choose an ordered grid \( F_1 = \{s_1^1,...,s_N^1\}  \subset (a,b) \).  
\item[]    Compute  \( J_1 = \{J(\{s_1^1\}),...,\ J(\{s_N^1\}) \} \).    
%}
\item[] Find $\Tilde{\textbf{k}}_1 = \underset{\textbf{k} \in J_1}{\operatorname{argmin}} f_1(\textbf{k})  $
\item[] Compute $\hat{\textbf{k}}_1$ solution of (\ref{P2}) with $p=1$ using  the Gauss-Newton algorithm with \( \Tilde{\textbf{k}}_1 \) as the starting point.
\item[] \( \hat{\boldsymbol{\tau}}_1 = J^{-1} (\hat{\textbf{k}}_1) \) 
\end{itemize}

\textbf{Forward addition} \\
\textbf{For} \(i=2,...,p\)

\begin{itemize}

\item[] Choose an ordered grid \( F_i = \{s_1^i,...,s_N^i\}  \subset (a,b) \).
\item[] Compute
\item[] \( J_i = \{ J \bigl(
 \hat{{\boldsymbol{\tau}}}_{j-1}\bigcup \{s_1^i\} \bigr),..., \bigl(
 \hat{{\boldsymbol{\tau}}}_{j-1}\bigcup \{s_N^i\} \bigr) \} \).

\item[] Find $\Tilde{\textbf{k}}_i = \underset{\textbf{k} \in J_i}{\operatorname{argmin}} f_i(\textbf{k})  $

\end{itemize}

\end{minipage}
}

\fbox{
\begin{minipage}{7cm}

\begin{itemize}

\item[] Compute $\hat{\textbf{k}}_i$ solution of (\ref{P2}) with $p=i$ using  the Gauss-Newton algorithm with \( \Tilde{\textbf{k}}_i \) as the starting point.
\item[] \( \hat{\boldsymbol{\tau}}_i = J^{-1} (\hat{\textbf{k}}_i) \) .

\end{itemize}
\textbf{EndFor}

\vspace{0.3cm}
\end{minipage}

}\\

Although there is no guarantee that this (or any other) algorithm will find the global minimizer of (\ref{P2}), we have found that it works well in practice. 
In our simulations and examples, knots have been added in the "right" order \cite{bib2},\cite{bib6}. This is important for the selection of the model, since the optimal number of knots \(p\) is never known in practice and will be chosen on the basis of sequences of intermediate nodes.

\section{Computational experiments}\label{sec4}

In this section, we present some computational results using our algorithm on functional data coming both on synthetic data and on data from a real-world application. More precisely, we present applications of three different clustering methods to evaluate the benefits of detecting clusters when using free knots splines estimation with two penalty terms with respect to free knots splines and free knots splines with one penalty term. To facilitate the notation, we will use \textit{FS0} to indicate the traditional free knots spline approximation, \textit{FS1} to indicate the free knots spline approximation with a single penalty term on the second derivative and \textit{FS2} to indicate the free knots spline approximation with a double penalty term.
\begin{figure}[!ht]
	\centering
	\includegraphics[width=5cm]{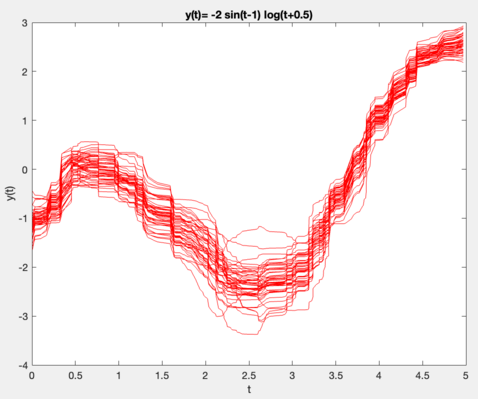}
	\quad\includegraphics[width=5cm]{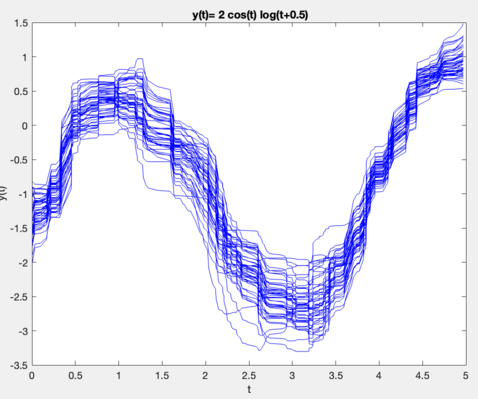}
	\quad\includegraphics[width=5cm]{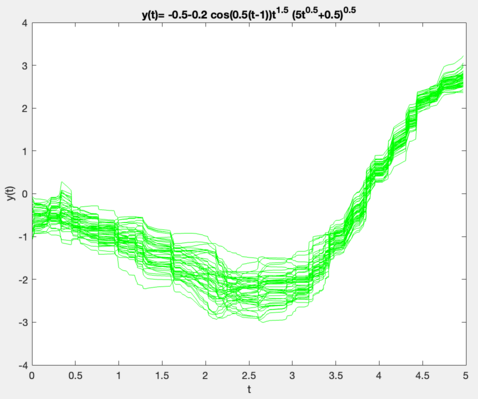}
	\quad\includegraphics[width=5cm]{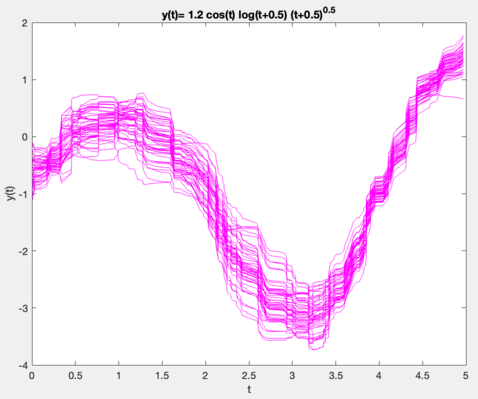}
	\caption{Simulated data.}
	\label{fig1}
\end{figure}

We will consider the classical k-means method for functional data \cite{bib7}, a model based clustering method \cite{bib8}
and a hierarchical agglomerative clustering methods 
\cite{bib12},\cite{bib13}, which we will refer to respectively R packages
as   \texttt{kmeans.fd}, \texttt{funFEM}, \texttt{fdahclust}.  %\texttt{fdakmeans},
\subsection{Synthetic datasets}
In the simulated scenario, four groups of 50 functions each were generated as in \cite{bib14} using the following functions as average functions
\begin{itemize}
\item 
       $ y(t)= -2 \sin(t-1)\log(t+0.5)$,
\item   $
        y(t)= 2\cos(t) \log(t+0.5)
   $,
   \item $ y(t)= -0.5-0.2\cos(0.5(t-1))t^{1.5} \sqrt{5 \sqrt{t}+0.5}$,
   \item $y(t)= 1.2 \cos(t) \log(t+0.5) \sqrt{t+0.5}$,
\end{itemize}
i.e. functions that resemble a negative sine (or cosine) wave with a regular oscillation period, multiplied either by the natural logarithm of \((t + 0.5)\) which modulates the amplitude  of the curve, causing it to progressively decrease as t increases, or 
the product of a combination  of log, power and square root functions which makes the curve strongly nonlinear and intricate.

Errors determined by a normal distribution with zero mean and standard deviation are added to each curve. Random errors are randomly generated and modeled to incorporate heterogeneity in the variance along the functional curves.
Each function was sampled in a common set of 50 randomly chosen points in the interval $[0,5]$.

Each group, as can be seen in Figure~\ref{fig1}, is characterized by differences in terms of amplitude, shape, and complexity. 

The first step in approximating the data was to determine the regularization parameters through cross validation: 
we chose the values \( \lambda_1 = 10^{-7}\) and \(\lambda_2 = 10^{-5}\) for \textit{FS2} obtained by minimizing with respect to $\lambda_1$ and $\lambda_2$ the GCV on a $L \times L$ grid, with \( L= \{10^{l }, \; l=-8,-6 \hdots 3,4 \} \) and \(\lambda_2 = 10^{-5}\) for \textit{FS1}.

\begin{figure}[!ht]
	\centering
	\includegraphics[width=5cm]{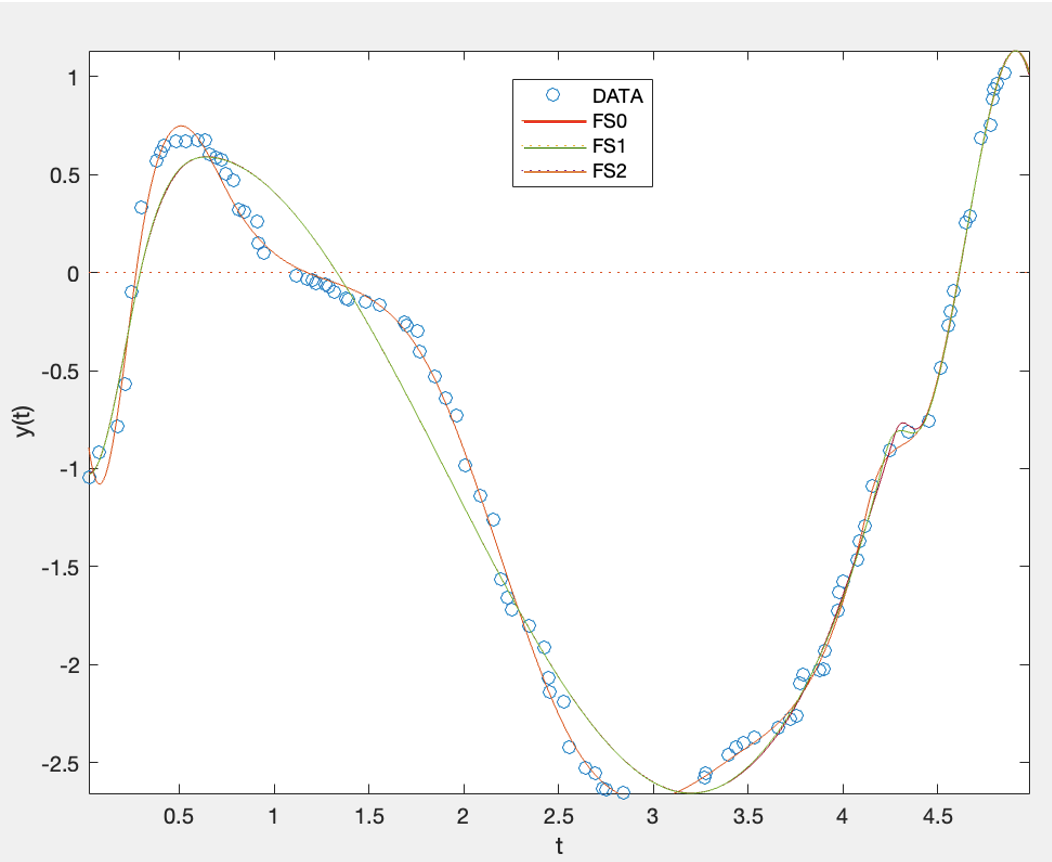}
	\quad\includegraphics[width=5cm]{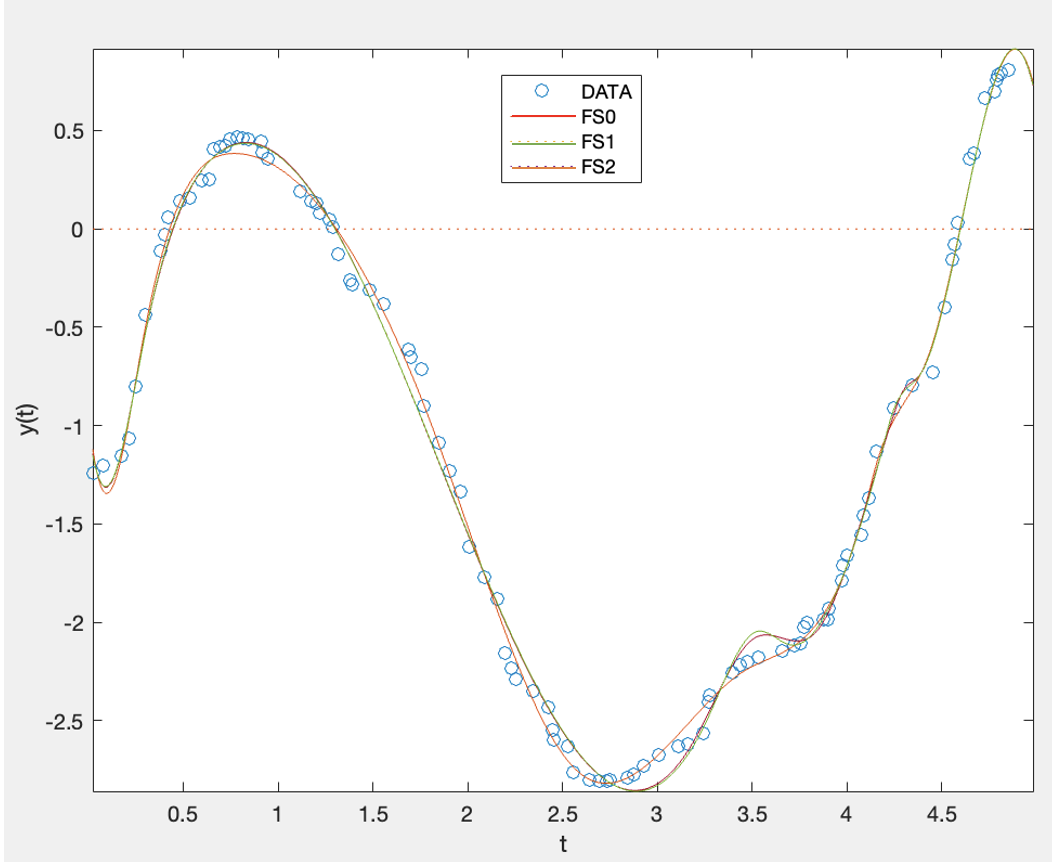}
	\includegraphics[width=5cm]{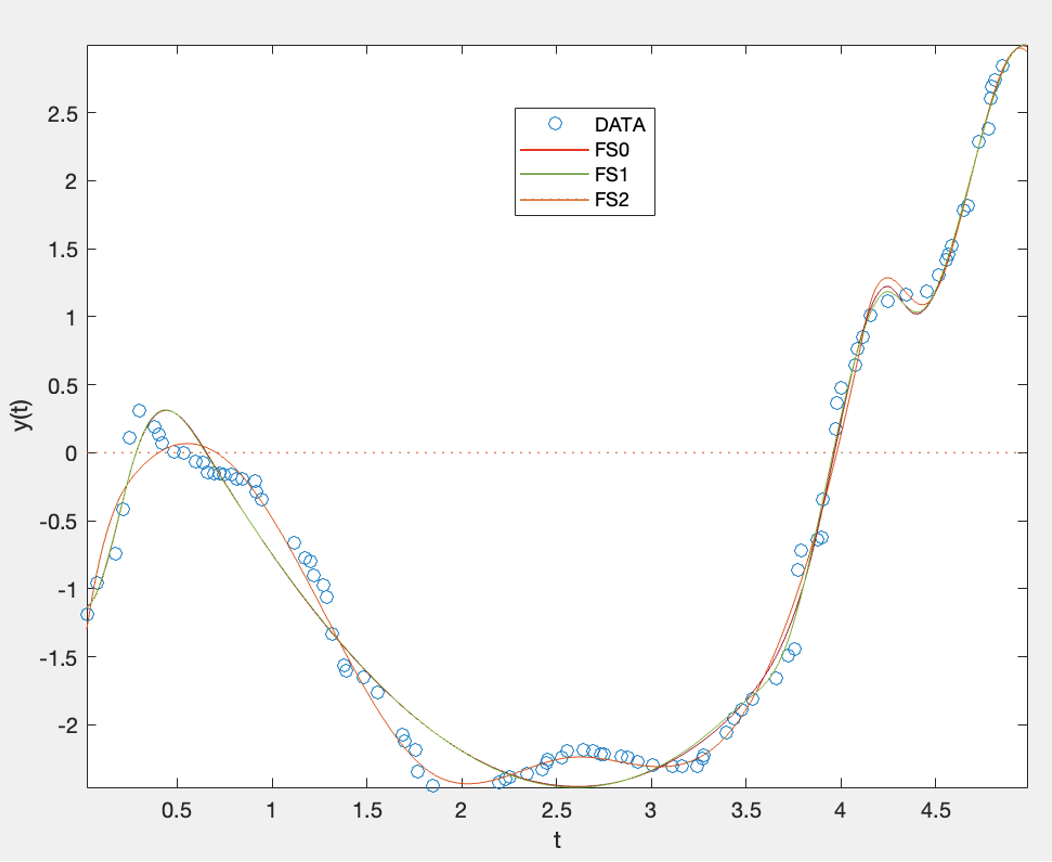}
	\quad\includegraphics[width=5cm]{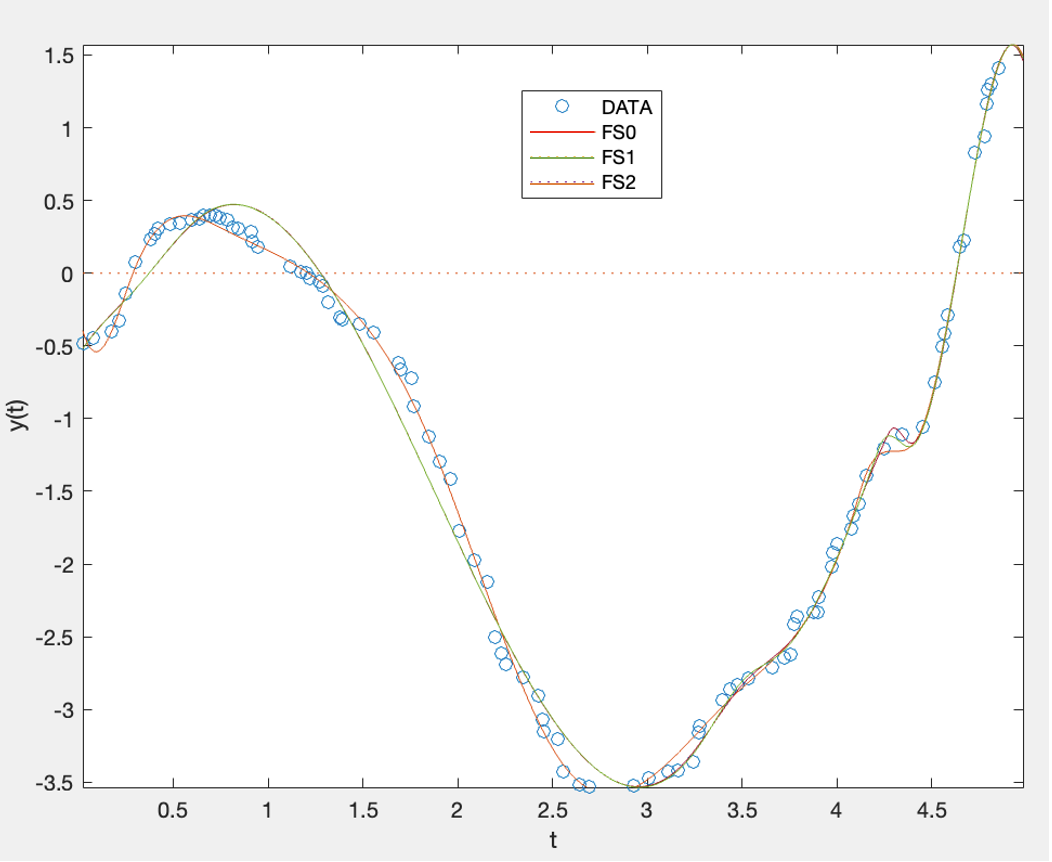}
	\caption{Comparison of \textit{FS0},\textit{FS1} and \textit{FS2} for data approximation in four case.}
	\label{fig2}
\end{figure}

The number of basis for the approximation, selected via cross validation, is set to 12. This simulation compares \textit{FS0}, \textit{FS1} and \textit{FS2}. The graphical results are shown in Figure~\ref{fig2}, and the numerical results are presented in Table~\ref{tab1}.

Performances of the two different approximation methods were measured using the classic Integrated Sum of Square Errors (ISSE) and its local version defined on the tails of the curves. The ISSE is a statistical metric used to assess the goodness of fit of a regression model or interpolation model to observed data. It is particularly useful in scenarios where the dependent variable is a continuous function of an independent variable, such as in functional data analysis. The ISSE is calculated by summing the squared differences between the observed values and the predicted values over the entire data range.
Similarly to the traditional ISSE, we define a Local ISSE (ISSE$_{inf}$,ISSE$_{sup}$) with the aim to quantify the fit in the initial and final portions of the curves. 

The expression of ISSE remains the same, but the boundaries of the integration interval need to be adjusted to reflect the specific regions of interest. This allows you to focus on the initial and final tails of the curve rather than the entire curve. The choice of regions for calculating the Integrated Sum of Squared Errors depends on the analysis objectives and the nature of the functional data or curves under examination. In our context, cross validation was used to test the model on different parts of the curves and identify the regions in the tails. Degrees of freedom ($df$), Integrated Sum of Square Errors, and the GCV scores are used to evaluate the quality of both the overall and local model fit.
The results in Table ~\ref{tab1} show that the lowest ISSE values were obtained using the free knots splines incorporating two regularization terms, both on the entire interval and on the tails.

\begin{table}[ht] \centering 
\caption{Table of comparison between \textit{FS0}, \textit{FS1} and \textit{FS2}.}
\label{tab1}

\begin{tabular}{@{}lcccc@{}}
\toprule

 \textbf{}	& \textbf{FS0}	& \textbf{FS1} & \textbf{FS2}\\
\midrule
$df$   & $0.120e+2$ & $0.119e+2$ & $0.119e+2$  \\
$ISSE $  & $0.154e+0 $ & $0.154e+0$ & $0.966e-1 $ \\
 \(ISSE_{inf}\)    & $0.977 e-1$ & $0.977 e-1 $ & $0.667 e-1 $\\
\(ISSE_{sup}\)   & $0.237 e-1 $ & $0.240e-1  $ & $0.227 e-1  $\\
$GCV$  & $0.506e+1$ & $0.509e+1$ & $0.198e+1$ \\
\bottomrule
\end{tabular}

\end{table}

The advantages obtained by using the dual penalty approach compared to free knots spline  are even more evident if we look at the clustering. 

\begin{figure}[!ht]
    \centering
    \includegraphics[width=8cm]{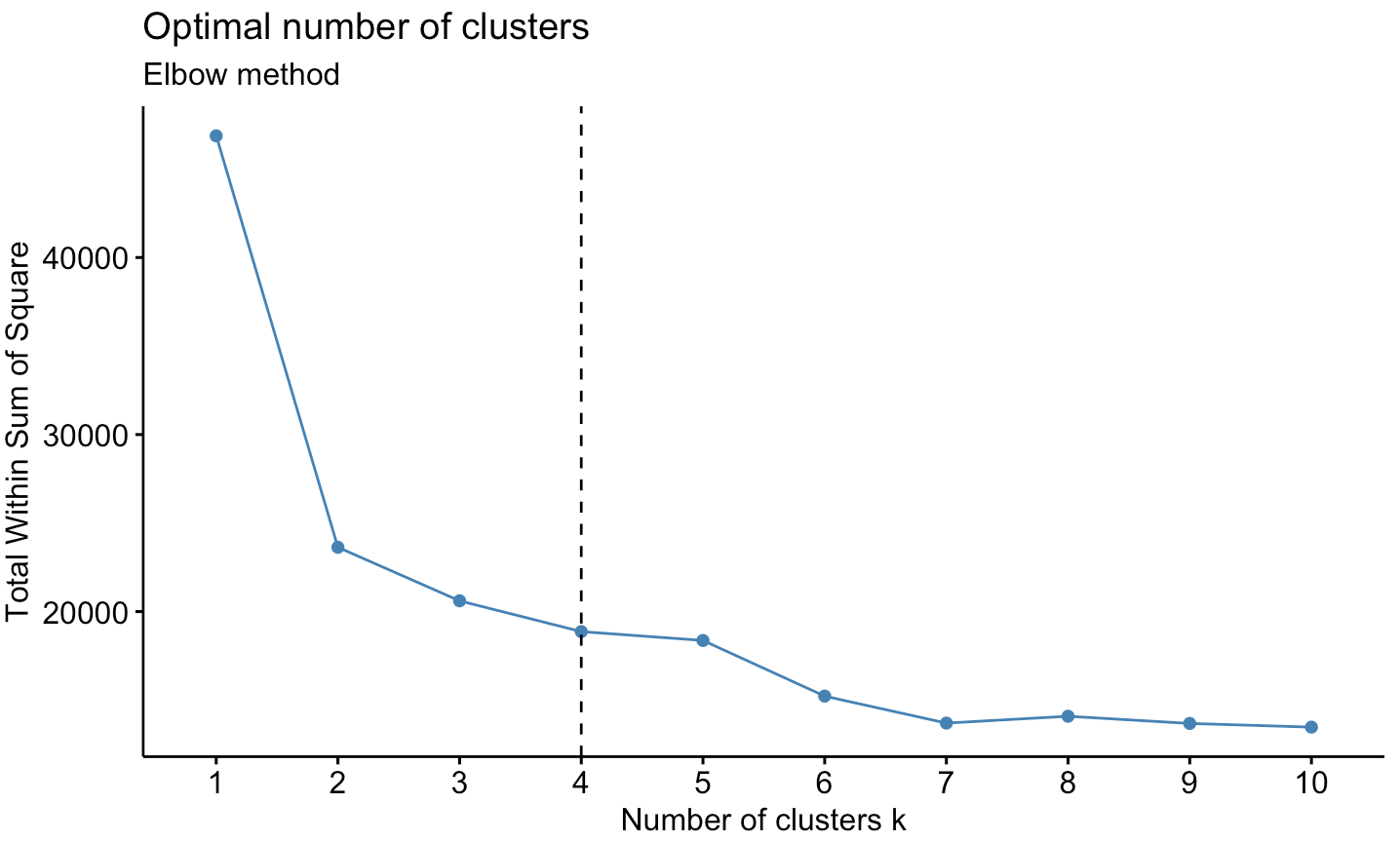}
    \caption{Evaluation of the
number of clusters in the simulation
dataset. The optimal number is
indicated from the dotted line.}
    \label{fig3}
\end{figure}

To determine the number of clusters, we used
 the Elbow method. In cluster analysis, the Elbow method is a heuristic used to identify the optimal number of clusters in a given dataset.
This method involves plotting the explained variation as a function of the number of clusters and selecting the 'elbow point' on the curve as the optimal number of clusters. The results in Figure~\ref{fig3} show that according to the index values, the optimal cluster number is c = 4. By applying \texttt{kmeans.fd} clustering to functional data, we can ascertain the number of curves assigned to each cluster for the methods under consideration.

\begin{table}[!ht] 
	\centering 
\caption{Table of comparison between clustering methods with \textit{FS0}, \textit{FS1} and \textit{FS2}: cluster cardinality and false positives ($FP$).}\label{tab2}

\begin{tabular}{@{}lcc|cc|cc@{}}
\toprule
%\textbf{}	& \textbf{Free knots spline}	& \textbf{FP}	&  \textbf{Free knots spline} & \textbf{FP}	&  \textbf{Free knots spline} & \textbf{FP}\\
 %\textbf{ }	& \textbf{$\lambda_1=\lambda_2=0$} & \textbf{}	& \textbf{$ \lambda_1=0, \lambda_2>0$ } & \textbf{} & \textbf{$\lambda_1>0, \lambda_2>0$} & \textbf{}\\ % Table header row
 \textbf{}	& \textbf{FS0}	& \textbf{FP}	&  \textbf{FS1} & \textbf{FP}	&  \textbf{FS2} & \textbf{FP}\\
\midrule
cluster 1	& 34 & 0 & 34 & 0 & 39 &  0\\
cluster 2    & 48 & 1 & 48 & 1 & 49 & 0\\
cluster 3    & 66 & 17 & 66 & 17 & 61 & 11\\
cluster 4   & 52 & 2 & 52 & 2 & 51 & 1\\
\bottomrule

\end{tabular}

\end{table}

\begin{figure}[!ht]
    \centering
    \includegraphics[width=7.7 cm]{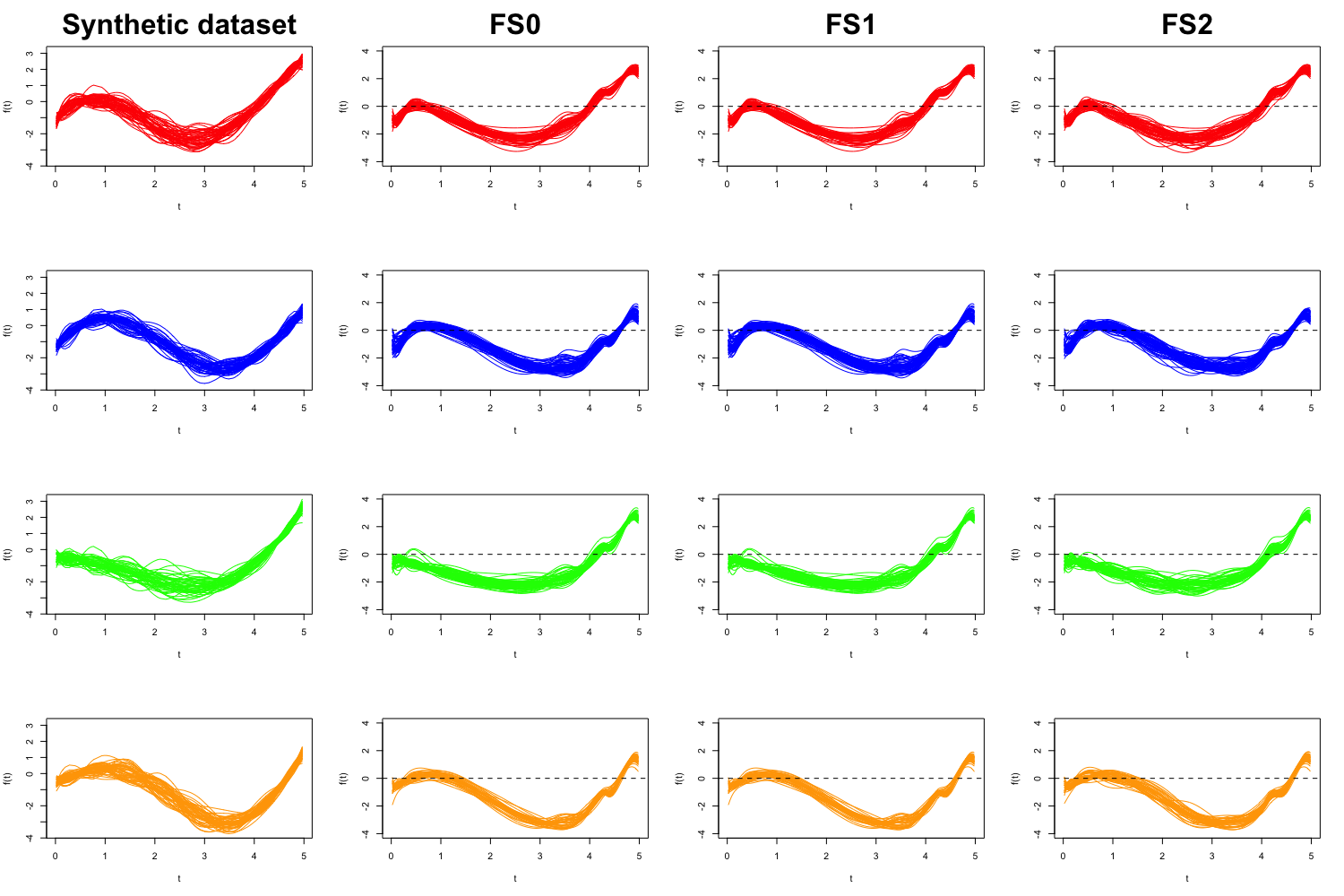}
    \caption{Clustering structure: Cluster 1 in red, Cluster 2 in blue, Cluster 3 in green, and Cluster 4 in orange. First column represents synthetic datasets,the second column represents the data approximated using \textit{FS0} segmented into clusters, the third column shows the data approximated using \textit{FS1} segmented into clusters, the last column illustrates the data approximated using \textit{FS2} segmented into clusters.}
    \label{fig4}
\end{figure}

Results reported in Table~\ref{tab2} and in Figure~\ref{fig4}, with respect to \textit{FS2},
many more misclassifications are observed for both \textit{FS0} and \textit{FS1}.

% For instance, in the case of $FS0$, the first cluster contains 92 curves, while the last cluster contains only 7 curves, whereas each cluster should ideally consist of 50 curves. 
The approximation that yields the best clusters is the one using \textit{FS2}.

In conclusion, it can be seen that introducing two regularization terms into the free knots spline approximation method led to have better approximation results also for further analyses such as clustering. Indeed, when data exhibit very similar shapes, clustering typically works regardless of approximation; the problem arises when dealing with data of highly dissimilar shapes, as seen in this case with simulated data, which a conventional method fails to capture. 

To consolidate the results obtained from clustering the curves approximated by the free knot spline, we demonstrate that no method of functional data clustering can yield the same or better results. From now on we no longer consider $FS1$ since just adding a regularization term does not improve the analysis through the use of clustering. 

\begin{table}[ht] \centering
\caption{Table of comparison between clustering methods with \textit{FS0} and \textit{FS2}: cluster cardinality and false positives ($FP$).}\label{tab3}%

\begin{tabular}{@{}lcc|cc@{}}
\toprule
%\textbf{Method} &
			%\textbf{Free knots spline }& \textbf{FP}	& \textbf{Free knots spline}	 & \textbf{FP}  \\
%\textbf{} &
			%\textbf{$\lambda_1=\lambda_2=0$ }& \textbf{}	& \textbf{$\lambda_1>0, \lambda_2>0$}	 & \textbf{}  \\
\textbf{Method} &
			\textbf{FS0}& \textbf{FP}	& \textbf{FS2}	 & \textbf{FP} \\
\midrule
$kmeans.fd$	& cluster 1 = 34 &  0 & cluster 1 = 39 & 0 \\
\textbf{} & cluster 2 = 48 &  1 & cluster 2 = 49 & 0\\
\textbf{} & cluster 3 = 66 & 17 & cluster 3 = 61 & 11 \\
\textbf{} & cluster 4 = 52 & 2 & cluster 4 = 51 & 1\\
\midrule
$funFEM  $  & cluster 1 = 56 &  41 & cluster 1 = 26 &  0 \\
\textbf{} & cluster 2 = 53 & 38 & cluster 2 = 17 & 0\\ 
\textbf{} & cluster 3 = 35 &  40 & cluster 3 = 100 & 50\\ 
\textbf{}	& cluster 4 = 56 & 43 & cluster 4 = 57 &  7\\ 
\midrule
%$fdakmeans$ & cluster 1 = 54 & 24 & cluster 1 = 37 & 0\\
%\textbf{} & cluster 2 = 74 & 32 & cluster 2 = 23 & 0 \\
%\textbf{} & cluster 3 = 24 & 0 & cluster 3 = 110	& 60\\ 
%\textbf{} & cluster 4 = 48 & 24 & cluster 4 = 30	&  0 \\ 

%\midrule
$fdahclust$  & cluster 1 = 69 & 19 & cluster 1 = 62 & 12 \\
\textbf{} & cluster 2 = 16 & 0 & cluster 2 = 8 & 0\\
\textbf{} & cluster 3 = 111 &  61 & cluster 3 = 87 & 37 \\ 
\textbf{} & cluster 4 = 4 & 24 & cluster 4 = 43 & 0 \\ 
\bottomrule
\end{tabular}
\end{table}

Results resented in Table~\ref{tab3} for all the clustering methods show how the best classification is achieved by applying clustering to curves approximated with $FS2$. Nevertheless, none of the classifications results are comparable to the classification obtained by the \texttt{kmeans.fd} method.

We can draw the same conclusions by evaluating the results obtained using the previous clustering methods by performing $150$ simulations. In order to compare the performances, we use the Adjusted Rand index $ARI$ measuring the percentage of times a clustering algorithm correctly determines the number of clusters out of a total of 150 simulations. Clustering can be thought of as a series of decisions where the goal is to group two individuals into the same cluster if and only if they are similar. A "true positive" ($TP$) decision correctly assigns two similar individuals to the same cluster, while a "true negative" ($TN$) decision correctly assigns two dissimilar individuals to different clusters. On the other hand, a "false positive" ($FP$) decision incorrectly assigns two dissimilar individuals to the same cluster, and a "false negative" ($FN$) decision incorrectly assigns two similar individuals to different clusters.

The Rand index (RI) is defined as:
 \[RI = \frac{TP + TN}{TP + FP + FN + TN}\]
and falls within the range between 0 and 1. A value of 0 indicates that the two data clusterings do not agree on any pair of points, while a value of 1 means that the data clusterings are identical. However, RI's value may not be close to 0 when category labels are randomly assigned, leading to potential issues. 

\begin{table}[ht] \centering
\caption{A comparison table of clustering methods, including \textit{FS0} and \textit{FS2}, based on 150 simulations. }\label{tab4}%

\begin{tabular}{@{}lcccc@{}}
\toprule
 \textbf{Method} & \textbf{} &
			\textbf{ARI FS0 } & \textbf{}  	& \textbf{ARI FS2}	 \\

\midrule
$kmeans.fd$  & \textbf{} & 0.78	 & \textbf{} & 0.83 \\
$funFEM $   & \textbf{}  & 0.54 & \textbf{} & 0.65 \\

$fdahclust$  & \textbf{}  & 0.52  & \textbf{}  & 0.64 \\
\bottomrule
\end{tabular}

\end{table}
To address this problem, the Adjusted Rand index (ARI) is introduced, defined as:
\[ARI = \frac{RI - E[RI]}{\max(RI) - E[RI]}.\]
ARI's range is between $-1$ and $1$, with values closer to $1$ indicating better clustering results.
From Table~\ref{tab4} we can observe that 
we can observe that for all the clustering methods, the use in the functional approximation
of $FS2$ leads to higher ARI
compared $FS0$.

\subsection{Application: Real Datasets }\label{sec5}

In this section, we aim to provide further confirmation of the validity and competitiveness of the proposed method by examining a real-world dataset.

We will examine the application of clustering to the COVID-19 dataset on “New cases in different countries from 2020 to 2021, specifically COVID-19 pandemic data for 30 countries. In this context, our goal is to demonstrate the feasibility of analyzing pandemic models through time series clustering using knot-free splines with two regularization terms for approximating functional data and finding taxonomies within the data. The data used in this paper were sourced from the World Health Organization (WHO) and the COVID-19 Data Hub\cite{bib14},\cite{bib15}. These databases are known for their transparency, open accessibility, and high credibility, ensuring a high level of accuracy.

As a significant number of countries experienced substantial outbreaks in March 2020, we selected this month as the starting point for our sequential data in this research. The endpoint of our dataset corresponds to the date of November 30st, 2021. As in \cite{bib40}, to mitigate the impact of factors such as varying population size, land area, population density, and population mobility across different countries, which can lead to significant differences in the magnitude of daily new COVID-19 cases,
we at first proceeded to standardize the raw data as follows:
%we have standardized the raw data for each country. The normalized variable for the number of daily new cases is as follows: 
\begin{equation*}
    y_{i,t}^* = \frac{y_{it}- \Bar{y}_{i}}{s_{i}} \;\;\;\;  t=1,2,...,T  \;\;\;\;  i = 1,2,...,N
\end{equation*}
where \( y_{i,t}^*\) represents the normalized number of daily new cases in the country \(i\) on day \(t\), \(y_{it}\) stands for the number of daily new cases in the country \(i\) on day \(t\), \(\Bar{y}_{i}\)  represents the mean of daily new cases in the country \(i\) during the whole study period, \(s_{i}\) is the standard deviation of daily new cases in the country \(i\) during the whole study period.

\begin{figure}[ht]
    \centering
    \includegraphics[width=8cm]{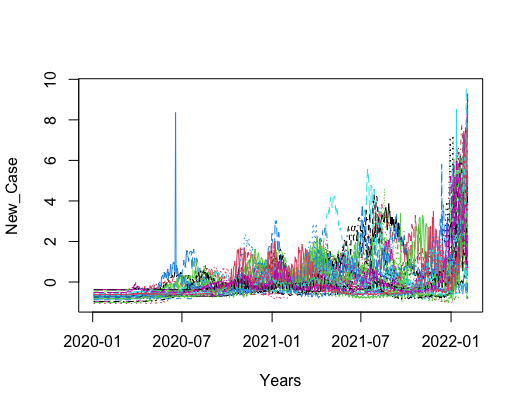}
    \caption{New Cases of Covid19 in different Country from 2020 to 2021. }
    \label{fig5}
\end{figure}

The data obtained can be seen in Figure~\ref{fig5}.
%The obtained data is visible in Figure~\ref{fig4}. 
The next step was to take the values \( \lambda_1= 10^{-3}\) and \(\lambda_2= 10^{-7}\). for the regularization parameters, obtained through a general cross-validation on a pre-established grid.
%The next step was to choose the regularization parameters through a general cross-validation process after building a grid for $\lambda_1$ and $\lambda_2$:

\begin{table}[ht] \centering
\caption{Numerical results.}
\label{tab5}%

\begin{tabular}{@{}lllll@{}}
\toprule
 \textbf{}	& \textbf{Free knots spline } & \textbf{Free knots spline } \\	
 \textbf{ }	& \textbf{$\lambda_1=\lambda_2=0$}	&  \textbf{$\lambda_1>0, \lambda_2>0$}\\
\midrule
$df$	& $0.230 e+2$ & $0.228e+2$  \\
$SSE$ & $0.330 e+0  $  & $0.312 e+0 $   \\
\(SSE_{inf}\)  & $0.722 e-2 $  &  $ 0.674e-2$ \\
\(SSE_{sup}\)  & $0.845 e-1 $  & $0.735 e-1$ \\
$GCV $ & $0.338e+1 $   &  $0.301 e+1$\\
\bottomrule
\end{tabular}

\end{table}

\begin{figure}[ht]
    \centering
    \includegraphics[width=6cm]{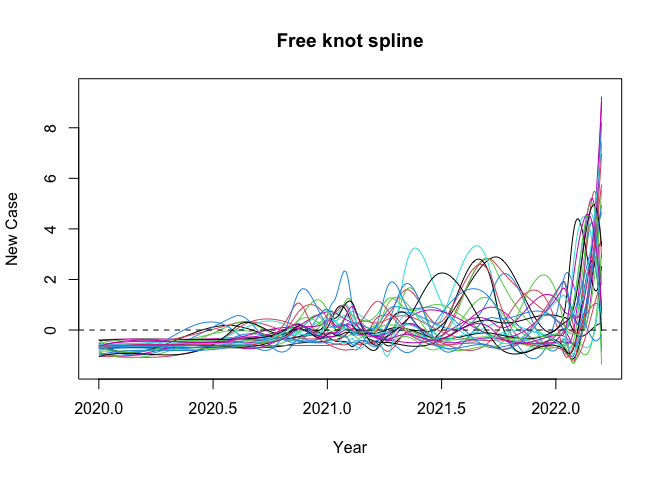}
    \quad \includegraphics[width=6cm]{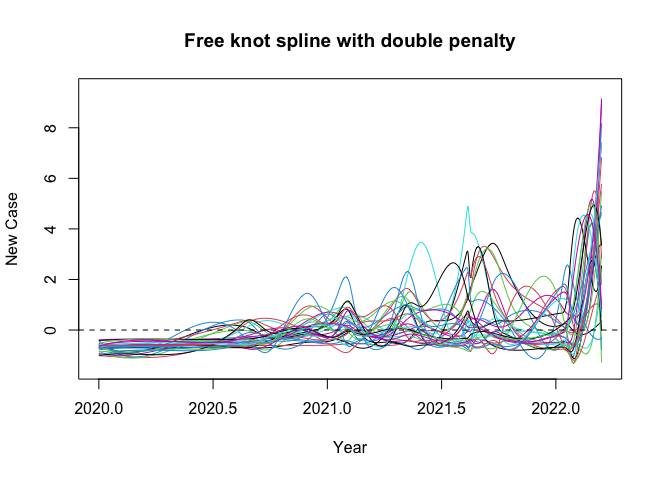}
    \caption{Covid19 Data smoothed by \textit{FS0} and \textit{FS2}.}
    \label{fig6}
\end{figure}

From Figure~\ref{fig6} and Table~\ref{tab5}, which provide comparisons between the two approximation methods, we can conclude that, as in previous cases, the double penalization approximation performs better along the extreme regions, ensuring that valuable information is not lost. 

\begin{figure}[ht]
    \centering
    \includegraphics[width=8cm]{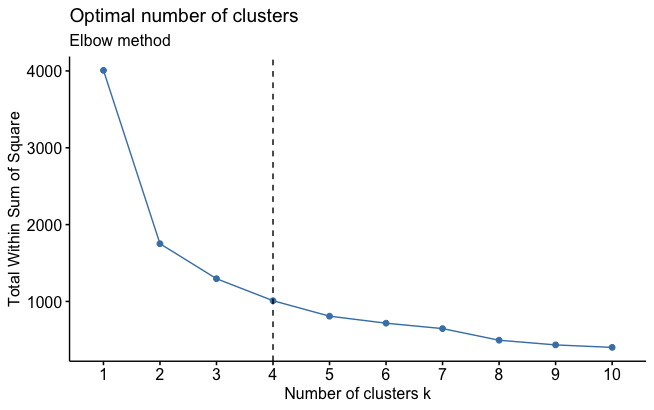}
    \caption{Evaluation of the
number of clusters in the simulation
dataset. The optimal number is
indicated from the dotted line.}
    \label{fig7}
\end{figure}

The advantages we gain from using the double penalization approximation method are observed in the application of clustering. To select the number of clusters, we apply the Elbow method. From the Figure~\ref{fig7}, we can determine that the appropriate number of clusters is 4.

The clustering method we will use is kmeans.fd. In the first analysis we perform is clustering on data approximated using \textit{FS2}. 

\begin{figure}[ht]
    \centering
    \includegraphics[width=8cm]{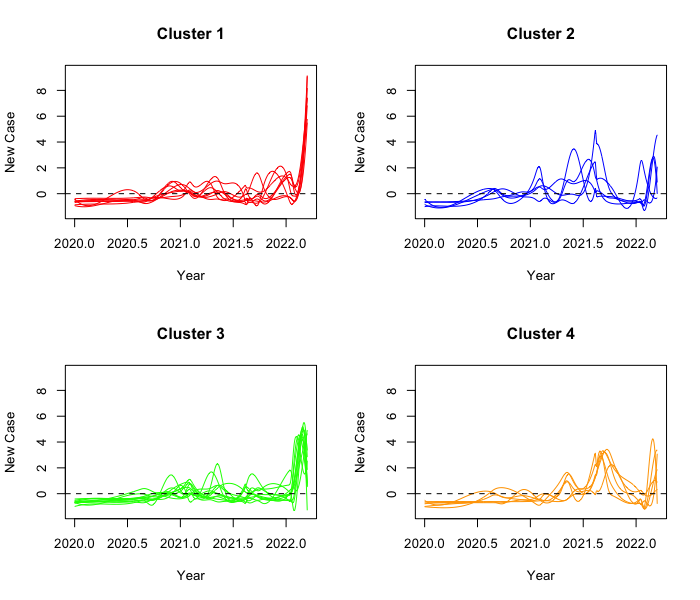}
    \quad\includegraphics[width=8cm]{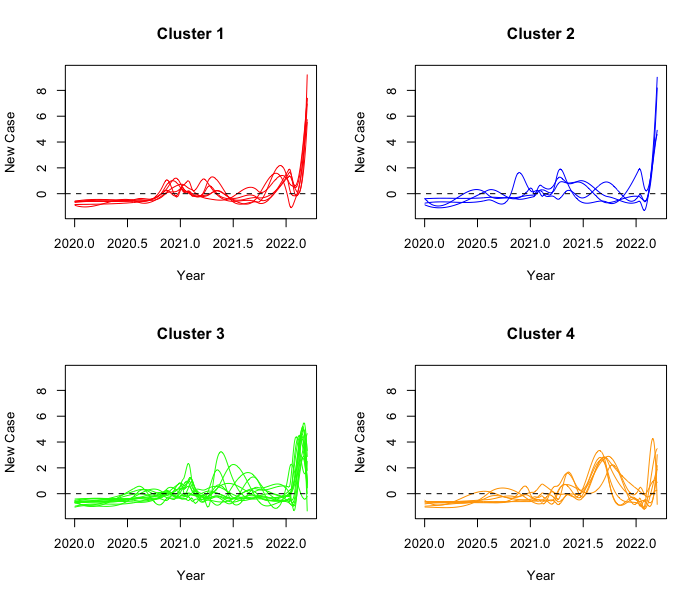}
    \caption{K-means clustering results by using \textit{FS2} (upper figure) and \textit{FS0} (down figure) }
    \label{fig8}
\end{figure}

Figure~\ref{fig8} displays the time series of daily new cases for each country based on the clusters. We can visually observe that the model varies significantly for countries in different clusters.

We can  observe that the pattern undergoes significant changes across countries within distinct clusters. The countries in Cluster 1 exhibit a consistent trend, followed by a sudden surge in cases from January 2022 to March 2022. Conversely, countries in Cluster 3 demonstrate a decline in new cases during the same period.

The countries in Cluster 2 show an increase in cases from the beginning of January 2021, followed by a decrease starting in February. During the same period, countries in Cluster 3 exhibited a consistent trend, followed by a rise in cases in March 2021. It can be observed that the pattern of pandemic development in Cluster 3 is often at the opposite pace to that of Cluster 4.

In general, each cluster of countries exhibits distinct characteristics that set their pandemic patterns apart from those of the other clusters.

\begin{table}[tbp]
\parbox{1\linewidth}{
\centering
\caption{Specific clustering results for daily new cases in 30 countries, in the case of \textit{FS2}. }\label{tab6}%

\begin{tabular}{@{}lllll@{}}
\toprule
 \textbf{Cluster}	& \textbf{Countries }	      \\	
\midrule
Cluster 1 &  Japan, Russian, Chile, Romania	 \\
\textbf{}  & Ukraine, Germany, Czech, Netherlands  \\
\midrule
Cluster 2 & Indonesia, South Africa, Brazil, India, Colombia \\
\midrule
Cluster 3  & Spain, Mexico, The U.S., The U.K.\\
\textbf{} &  Argentina, Turkiye, France, Canada \\ 
\textbf{} &  Italy, Poland, Peru \\   
                  \midrule
Cluster 4  & Philippines, Malaysia,  Thailand    \\
\textbf{} & The Islamic Republic of Iran, Bangladesh, Iraq     \\ 
\bottomrule
\end{tabular}

}
\end{table}

\begin{table}[tbp]
\parbox{1\linewidth}{
\centering
\caption{Specific clustering results for daily new cases in 30 countries, in the case of \textit{FS0}. }\label{tab7}%

\begin{tabular}{@{}lllll@{}}
\toprule
 \textbf{Cluster}	& \textbf{Countries }	      \\	
\midrule
Cluster 1  &    Russia, Romania, Ukraine  \\
\textbf{} &   Netherlands, Germany, Czech  \\
\midrule
Cluster 2  &  Japan, Brazil, Chile, Poland\\

\midrule
Cluster 3 &  Mexico, South Africa, The U.S., The U.K.\\
\textbf{} & Peru, India, Argentina, Colombia \\  
\textbf{}  & Canada, Turkiye, Italy, France, Spain \\
 
                  \midrule
Cluster 4 &  Malaysia, Thailand, The Islamic Republic Iran, \\
\textbf{} &  Indonesia, Bangladesh, Iraq, Philippines	 \\\bottomrule
\end{tabular}

}
\end{table}

Table~\ref{tab6} shows the specific clustering results for the 30 countries. We can see that all Cluster 1 and Cluster 4 countries are located in regions of Asia, while most of Cluster 3  is located in different regions in the European region.

For this reason we might think that geographical location can influence clustering. However, geographical proximity is not a decisive factor for clustering; just think of Cluster 2.

All the considerations we have made so far concern clustering on data approximated using \textit{FS2}. From the Figure~\ref{fig8} (upper figure) and the Table~\ref{tab6}, we can analyze the results of clustering on data approximated using \textit{FS0}. The Figure~\ref{fig8} (down figure) and Table~\ref{tab7} displays the different clusters for \textit{FS0}: the clusters are not well-defined, and the curves, not being correctly allocated to the clusters, making it challenging to read and interpret the data. The considerations that were previously made in this case are not as clear: in Clusters 1 and 2, we observe a similar trend, especially towards the end of 2021. Similar observations can be made for Clusters 3 and 4, which exhibit similar trends. Therefore, in the case of $FS0$, specific characteristics to characterize each cluster are not found.

\section{Conclusions}\label{sec6}

 In this work, we considered the use of free knots spline in the context of functional data estimation, analyzing, in particular, the impact of the roughness regularization term. More specifically, we compared two different penalty regularization schemes, namely a standard regularization scheme with a one parameter roughness term handling the boundedness of the variability of the function \cite{bib4} and a two-penalty regularization scheme which controls monotonicity  and smoothness. Our numerical experiments seem to demonstrate that compared to the  free knots spline without any penalty, things do not change much when using the one-parameter scheme, while the two-parameter scheme shows notable improvements. However, the most significant advantages appear in the data analysis phase based on the functional data approximation obtained.  In particular, when applied to simulated data, our method shows a higher improved ability to detect ties, highlighting a clearer clustering structure.
 A promising direction emerges from the non-linear effects linked to the selection of knots in the initial basis. The incorporation of machine learning method to enhance the analytical approach, open a different perspective for continuous advancements in the domain of functional data analysis.

\backmatter

\bmhead{Acknowledgements}
This work was partially supported by Italian Ministry of University and Research (MIUR), PRIN 2022 Project: \emph{Numerical Optimization with Adaptive Accuracy and Applications to Machine Learning
}, grant n. 2022N3ZNAX, PNRR PRIN 2022 Project: \emph{A multidisciplinary approach to evaluate ecosystems resilience under climate change}, grant n. P2022WC2ZZ, and 
by Istituto Nazionale di Alta Matematica -
Gruppo Nazionale per il Calcolo Scientifico (INdAM-GNCS), by MIUR.

%\printbibliography
\medskip
%\bibliography{ref.bib}
\bibliographystyle{plainnat}
\bibliography{refn.bib}

\end{document}